%% file: scisoc2026.tex
\documentclass[sigconf,nonacm]{acmart}

\usepackage{booktabs}
\usepackage{graphicx}
\usepackage{amsmath}
\usepackage{hyperref}
\AtBeginDocument{%
  }

\acmYear{2026}
\acmDOI{10.5281/zenodo.20954144}
\acmConference[KDD 2026 Workshop]{Accepted for presentation at the ACM KDD2026 SciSoc Workshop 2026 (non-archival).}{Aug 09 - 13,
  2026}{Jeju, Korea}
\begin{document}

\title{LLMoxie: Exploring Agentic AI for Scientific Software Development}

\author{Landung Setiawan}
\affiliation{
  \institution{eScience Institute}
  \institution{University of Washington}
  \city{Seattle}
  \state{Washington}
  \country{USA}
}
\email{landungs@uw.edu}

\author{Anant Mittal}
\affiliation{
  \institution{eScience Institute}
  \institution{University of Washington}
  \city{Seattle}
  \state{Washington}
  \country{USA}
}
\email{anmittal@uw.edu}

\author{Cordero Core}
\affiliation{
  \institution{eScience Institute}
  \institution{University of Washington}
  \city{Seattle}
  \state{Washington}
  \country{USA}
}
\email{cdcore@uw.edu}

\author{Anshul Tambay}
\affiliation{
  \institution{eScience Institute}
  \institution{University of Washington}
  \city{Seattle}
  \state{Washington}
  \country{USA}
}
\email{anshul37@uw.edu}

\author{Carlos Garcia Jurado Suarez}
\affiliation{
  \institution{eScience Institute}
  \institution{University of Washington}
  \city{Seattle}
  \state{Washington}
  \country{USA}
}
\email{carlosg@uw.edu}

\author{David A. C. Beck}
\affiliation{
  \institution{eScience Institute}
  \institution{Dept of Chemical Engineering}
  \institution{University of Washington}
  \city{Seattle}
  \state{Washington}
  \country{USA}
}
\email{dacb@uw.edu}

\author{Andrew J. Connolly}
\affiliation{
  \institution{eScience Institute}
  \institution{Dept of Astronomy}
  \institution{University of Washington}
  \city{Seattle}
  \state{Washington}
  \country{USA}
}
\email{ajc@astro.washington.edu}

\author{Vani Mandava}
\affiliation{
  \institution{eScience Institute}
  \institution{University of Washington}
  \city{Seattle}
  \state{Washington}
  \country{USA}
}
\email{vani1@uw.edu}


\hyphenpenalty=10000
\tolerance=2000
\emergencystretch=10pt

\raggedbottom
\begin{abstract}
  \input{content/0_abstract.tex}
\end{abstract}

\begin{teaserfigure}
  \includegraphics[width=\columnwidth, scale=0.5]{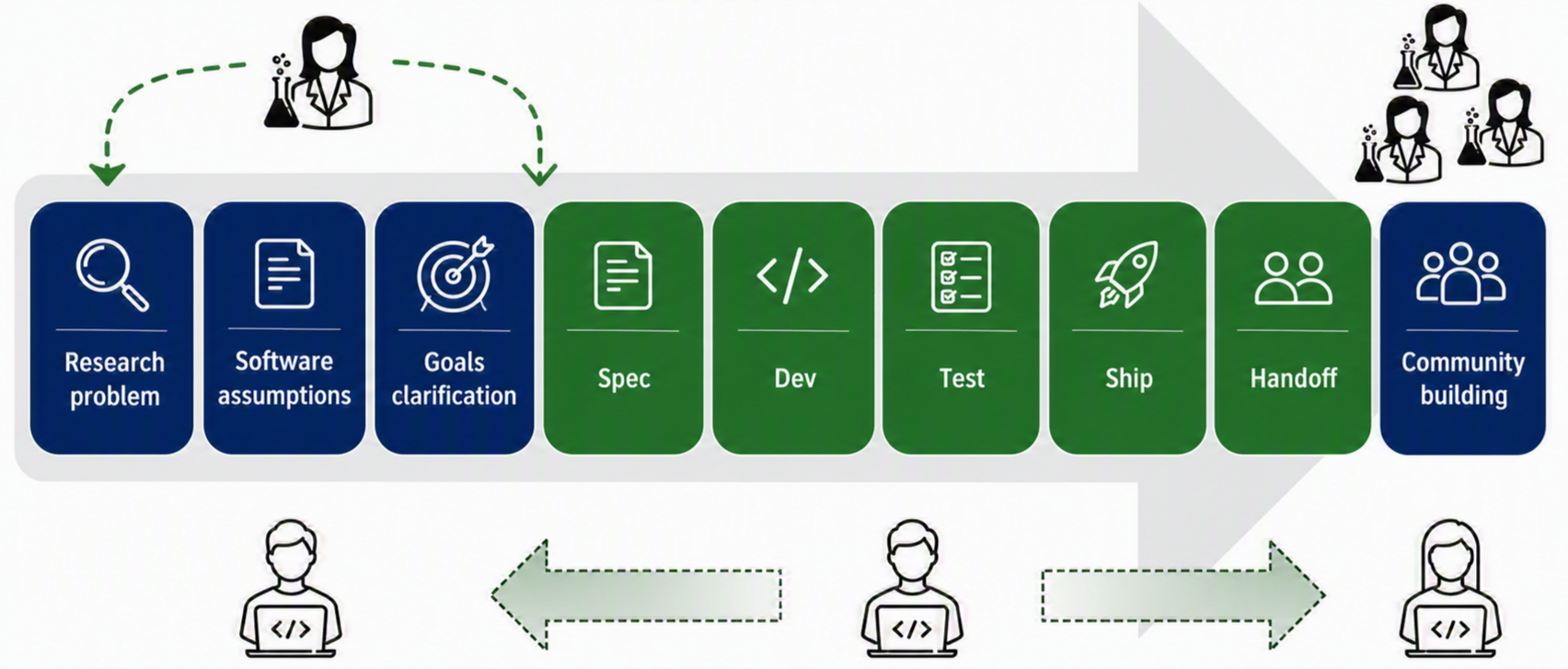}
  \caption{An RSE-oriented scientific software development lifecycle. Conventional engineering stages (Spec, Dev, Test, Ship) are bracketed by science-specific phases: upstream framing of the research problem, software assumptions, and goals clarification, and downstream handoff and community building. Scientists drive the bookend phases while research software engineers carry the work across the middle, illustrating where AI-assisted coding agents must integrate to support scientific practice.}
  \Description{A horizontal pipeline of nine stages inside an arrow. Three blue stages on the left (Research problem, Software assumptions, Goals clarification) and two blue stages on the right (Handoff, Community building) flank four green engineering stages in the middle (Spec, Dev, Test, Ship). A scientist icon sits above the left and right blue stages; developer icons sit below the engineering stages, indicating that scientists lead the framing and community phases while RSEs lead specification through shipping.}
  \label{fig:teaser}
\end{teaserfigure}

\maketitle

\section{Introduction}
\label{sec:introduction}
\input{content/1_introduction}

\section{Background \& Motivation}
\label{sec:background}
\input{content/2_ai_adoption_rse}

\section{System Architecture}
\input{content/3_system_architecture}
 
\section{RSE-Plugins Architecture and Design}
\label{sec:rse_plugins}
\input{content/4_rse_plugins}

\section{Research Capabilities and Operational Modes}
\label{sec:capabilities}
\input{content/5_research_capabilities_and_operational_modes} 

\section{Implications and Future Work}
\label{sec:implications}
\input{content/6_implications_and_future_work}

\section{Limitations}
\label{sec:limitations}
\input{content/7_limitations}

\section{Conclusion}
\label{sec:conclusion}
\input{content/8_conclusion}

\section{Acknowledgments}
\label{sec:acknowledgments}
\input{content/9_acknowledgments} 
\bibliographystyle{ACM-Reference-Format}
\bibliography{uw-ssec-new}

\end{document}

%% file: content/0_abstract.tex
In this paper, we describe \emph{LLMoxie}, an institutional AI platform whose three-tiered architecture supports multi-cloud and on-premise inference, a LiteLLM/MLflow control plane for authentication, budgeting, PII masking, and observability, and an application augmentation layer for AI coding agents. Layered on top, an open-source \emph{RSE-Plugins} ecosystem encodes accumulated RSE knowledge as a Plugin-Agent-Skill hierarchy spanning scientific Python practice, domain-specific knowledge, a six-phase research-and-implement workflow, and project lifecycle management. Scientific software is judged less by raw code quality than by whether it can be cited, audited, reproduced, and extended. Off-the-shelf AI coding agents, optimized against commercial software benchmarks, are poorly calibrated for this setting: they ignore the conventions of the scientific Python libraries they invoke, mishandle sensitive or embargoed data, and leave decision trails that are difficult to reconstruct after the fact. We report on twenty months of practice at a university-based research software engineering (RSE) center, where RSEs embedded across astronomy, earth and climate science, agriculture, and health projects worked to close this gap. We characterize the recurring infrastructure, governance, and process challenges of adopting Agentic AI inside a multi-domain RSE center, describe the platform and plugin design, and distill operational lessons from real scientific software deployments. Together, the platform and plugins shift AI coding agents from generic code generators into domain-aware collaborators that respect community norms and produce auditable provenance of technical reasoning. 

%% file: content/1_introduction.tex
Our research software engineering center, Scientific Software Engineering Center, is a professional software engineering organization embedded within the University of Washington, Seattle. It serves as a campus-wide collaborator for domain scientists, translating research questions into durable software systems by clarifying scientific goals, designing reproducible workflows, testing underlying assumptions, and delivering maintainable tools. Since its establishment, the center has delivered more than 20 multi-institution scientific software projects, convened workshops and events reaching over 1,000 participants, sustained a graduate scholars program, and partnered with principal investigators across 32 organizations worldwide. Situated within University of Washington's  data-science institute, i.e., eScience Institute whose mission centers on data-intensive science and responsible AI practice, the center's research software engineers (RSEs) work alongside research scientists and data scientists to deliver software that satisfies both engineering quality bars and the standards of scientific practice.

That dual mandate of production-quality engineering and scientific reproducibility is what makes scientific software a distinctive setting for AI-assisted development. Within professional communities that work at the intersection of software and science, there is growing recognition that the RSE profession is positioned to become central to the responsible integration of AI into research~\cite{gesingRSEs2035Surviving2025, mcinnesReport2025Workshop2025, hughesAI4ScienceEmpowerFifth2022}. Unlike most commercial software, research code is judged less by its raw quality and robustness than by whether it can be cited, audited, reproduced, and extended by other scientists, often years after the original authors have moved on~\cite{hunter-zinckTenSimpleRules2021}. It typically operates over specialized data formats (e.g.,~\texttt{FITS}, \texttt{NetCDF}, \texttt{Zarr}, \texttt{HDF5}), relies on a tightly coupled scientific Python stack, runs on heterogeneous infrastructure ranging from laptops to high-performance computing clusters, and is written by teams in which professional software engineers are the exception rather than the rule. The sociotechnical landscape of scientific research software is complex and was not designed for engineering resilience or governance~\cite{hocquetnatcom2024, mcinnesReport2025Workshop2025}. Contemporary AI coding assistants, whose training data and evaluation benchmarks are dominated by commercial software development, are poorly calibrated and under-studied for research software. Specialized scientific-coding benchmarks~\cite{ainsteinbench2025, tianSciCodeResearchCoding2024} have emerged in response, yet even frontier models perform poorly on them and their findings are not systematically incorporated back into frontier model development.

While postdoctoral researchers report using generative AI tools for a range of research tasks~\cite{nordlingHowChatGPTTransforming2023, obrienHowScientistsUse2025, besserHowGenerativeAI2026}, AI-assisted research software engineering specifically remains an emergent field~\cite{farshidiAdvancingResearchSoftware2025, bridgefordTenSimpleRules2025}. A summer 2025 study of generative AI adoption among scientists who write code~\cite{obrienSurveyGenerativeAI2026} found that respondents overwhelmingly prefer general-purpose, proprietary conversational tools such as ChatGPT accessed through web browsers over developer-specific tools integrated into programming environments. A second 2025 study of more than 1{,}500 research software repositories~\cite{farshidiAdvancingResearchSoftware2025} found that only 17\% exhibited both high AI usage and high software engineering maturity. These patterns have raised concern that over-reliance on general-purpose conversational models places scientific code at risk~\cite{obrienThreatsScientificSoftware2025, bridgefordTenSimpleRules2025, kabirStackOverflowObsolete2024}, with downstream consequences for open-science integrity and reproducibility~\cite{hosseiniOpenScienceGenerative2024}.

From September 2024 to May 2026, our center engaged directly with this landscape to close the widening gap between AI-led software creation in industry and scientific software creation in academic research environments. Early efforts framed the problem as one of context: Retrieval-Augmented Generation (RAG) over scientific literature and pre-publication data \cite{zotero-item-5515} enabled general-purpose models to speak credibly about a given research domain, though even long-context RAG pipelines exhibit characteristic failure modes as corpora and contexts grow~\cite{LongContextRAGMon08/12/2024-12:46}. As frontier models improved and tool-using agents matured around the Model Context Protocol (MCP)~\cite{ganRAGMCPMitigatingPrompt2025, belcakSmallLanguageModels2025}, the binding constraint shifted from raw model capability to the absence of structured, domain-aware context and process scaffolding around the model. Off-the-shelf coding agents produce running Python while ignoring the conventions of the communities whose libraries they invoke~\cite{korenVibeCodingKills2026, bridgefordTenSimpleRules2025}, mishandle sensitive or embargoed data~\cite{hosseiniOpenScienceGenerative2024}, and leave decision trails that are difficult to reconstruct after the fact~\cite{storeyTechnicalDebtCognitive2026}. This is a particularly poor fit for scientific work, where the reasoning process is itself part of the artifact. Recent evaluations of agentic systems on realistic tasks reach a consistent conclusion: benchmark performance routinely overstates how reliably and auditably these systems behave in deployment~\cite{kapoorHolisticAgentLeaderboard2025, labanLLMsGetLost2025}.

RSE centers are unusually well positioned to address this gap~\cite{gesingRSEs2035Surviving2025, mcinnesReport2025Workshop2025}. Embedded across many scientific domains simultaneously, an RSE center observes the same failure modes recur across geoscience, astronomy, climate, agriculture, and health projects, and can therefore design a response that generalizes rather than overfits to any single laboratory. The substance of that response is to encode accumulated RSE knowledge into reusable context that AI agents can consume: packaging discipline, testing regimes for numerical code, documentation practice, reproducibility tooling, and structured research workflows~\cite{hunter-zinckTenSimpleRules2021, bridgefordTenSimpleRules2025}.

This paper reports on the result: \textbf{LLMoxie}, an institutional AI platform, together with \textbf{RSE-Plugins}, a hierarchical ecosystem of Claude-compatible plugins, agents, and skills that encode RSE practice as composable context for AI coding agents. LLMoxie provides a governed, multi-cloud inference layer with authentication, budgeting, PII masking, and observability; RSE-Plugins layer scientific Python conventions, domain-specific knowledge, and a structured research-and-implement workflow on top of a coding agent (Claude Code in our deployments). Together, they reposition the LLM from a generic code generator into a domain-aware collaborator that respects community norms and emits auditable trails of technical reasoning. 
Here we describe:
\begin{itemize}
    \item A characterization of the recurring infrastructure, governance, and process challenges of adopting agentic AI inside an RSE center serving multiple scientific domains, drawn from a portfolio of active projects.
    \item The design of LLMoxie, a three-tiered platform that separates inference, governance, and application-level augmentation, with rationale for each layer (Section~\ref{sec:architecture}).
    \item RSE-Plugins, a Plugin-Agent-Skill hierarchy that encodes scientific Python practice, domain-specific knowledge, and a six-phase research-and-implement workflow as reusable context for coding agents (Section~\ref{sec:rse_plugins}).
    \item Operational lessons from deploying this stack on real scientific software projects (Section~\ref{sec:implications}).
\end{itemize}

Unlike generic model-routing and inference optimization AI infrastructure tools~\cite{liLLMRouterBenchMassiveBenchmark2026}, LLMoxie shapes a broader research AI orchestration platform focused on integrating retrieval systems, scientific tools, scientific research knowledge bases, and extensible plugins into coherent research workflows. Its capabilities include tools, datasets, workflows, and execution environments with the LLM acting as only one component inside a larger research oriented ecosystem.

The remainder of the paper traces our adoption path from RAG-era experiments to the current LLMoxie design (Section~\ref{sec:background}), presents the three-tier platform and the RSE-Plugins hierarchy (Sections~\ref{sec:architecture} and~\ref{sec:rse_plugins}), surveys capability surfaces, production deployment, and engineering practices on active projects (Section~\ref{sec:capabilities}), and closes with implications, limitations, and conclusions (Sections~\ref{sec:implications},~\ref{sec:limitations}, and~\ref{sec:conclusion}).

%% file: content/2_ai_adoption_rse.tex

%
Between September 2024 and May 2026, our center engaged with the AI-assisted RSE landscape through a portfolio of active scientific software projects. Our traditional delivery model---six-to-nine-month engagements with science principal investigators (PIs) covering requirements gathering, specification, implementation, code review, testing, documentation, and packaging---provided a structured baseline against which to assess where generative AI tooling helped, where it failed, and what institutional scaffolding was missing.

The first concrete pilot embedded scientific literature and pre-publication data into a vector database and exposed it to general-purpose models through Retrieval-Augmented Generation (RAG), in collaboration with researchers at a large astronomical survey project \cite{zotero-item-5515}. Consistent with our open-science mandate, the initial deployment used only open-source models olmo \cite{Groeneveld2024OLMoAT}. Once the survey data left embargo, open-weight models served through community runtimes (e.g.,~\texttt{ollama}) reached state-of-the-art performance on the survey's question-answering tasks, eliminating the RAG advantage for already-public corpora. RAG remained valuable, however, for sensitive and pre-publication data where the underlying corpus could not be shipped to a hosted model.

This first wave clarified that the binding constraint was not model quality but the absence of governed, science-aware infrastructure surrounding it. In mid-2024 our team was awarded federally-funded national AI compute resources \cite{llmaven2026nairr} to build domain-agnostic AI infrastructure for scientists; the original proposal centered on hosted inference plus RAG. Over the following year, the release of the Model Context Protocol (MCP) and the rapid maturation of agent- and skill-based tooling shifted the locus of context engineering from retrieval over a single corpus to composable tools, agents, and skills coordinated by a coding agent. In response to changing landscapes, we evolved our effort to an agent- and skill-centric platform rather than a RAG service. Section~\ref{sec:architecture} describes the resulting platform.

%% file: content/3_system_architecture.tex
\label{sec:architecture}
 
\begin{figure}
    \centering
    \includegraphics[width=\columnwidth]{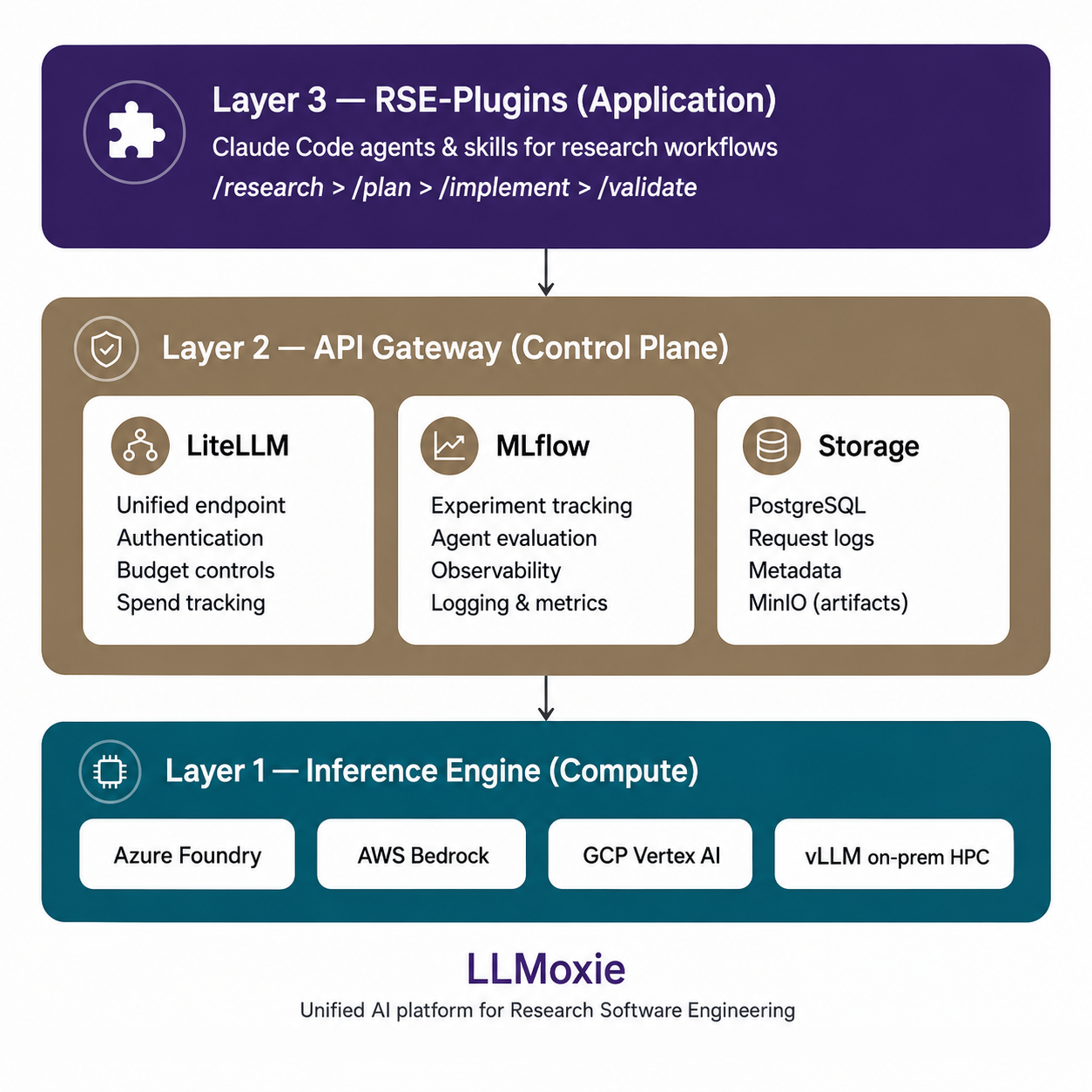}
    \caption{LLMoxie's three-tiered architecture for AI-enabled research software engineering. The inference layer routes requests across cloud (Azure Foundry, AWS Bedrock, GCP Vertex AI) and on-premise (vLLM/HPC) backends; the LiteLLM/MLflow control plane provides a unified OpenAI-compatible gateway with authentication, rate limiting, budget enforcement, PII masking, and observability; and the application augmentation layer delivers domain-specific workflows to Claude Code via the RSE-Plugins ecosystem.}
    \Description{A layered diagram of the LLMoxie platform showing three horizontal tiers. The bottom tier is the inference layer, with boxes for Azure Foundry Models, Amazon Bedrock, Google Vertex AI, and on-premise vLLM on HPC. The middle tier is a control plane built on LiteLLM and MLflow, exposing a single OpenAI-compatible endpoint and providing authentication, rate limiting, per-user and per-team budgets, PII masking via Microsoft Presidio, request logging to PostgreSQL, and model evaluation and observability. The top tier is the application augmentation layer, where Claude Code consumes the RSE-Plugins ecosystem organized as a Plugin-Agent-Skill hierarchy used by research software engineers.}
    \label{fig:llmoxie}
\end{figure}

LLMoxie implements a three-tiered architectural framework that separates the model-serving inference layer, the control plane for access control and governance, and the application-level agents and skills, as illustrated in Figure~\ref{fig:llmoxie}.

\subsection{Inference Layer}
\label{subsec:inference}
 
The foundational tier abstracts the underlying computational substrate, permitting transparent routing across heterogeneous inference backends. The system supports cloud-native deployment on Azure (via Microsoft Foundry Models), Amazon Web Services (via Amazon Bedrock), and Google Cloud Platform (via Vertex AI), as well as on-premise deployment via vLLM in high-performance computing (HPC) environments. This flexibility allows institutional deployment decisions to track existing infrastructure investments, data governance requirements, and cost considerations, including support for credit-based cloud subscriptions.
 
\subsection{Control Plane and Governance Layer}
\label{subsec:control}
 
An intermediate gateway layer, implemented with LiteLLM and MLflow, provides unified API access to the inference backends through a single OpenAI-compatible endpoint. This abstraction permits transparent provider switching without application-level modifications. The control plane implements the governance mechanisms listed below:
 
\begin{itemize}
    \item Authentication and authorization
    \item Request rate limiting (requests per minute and tokens per minute)
    \item Per-user and per-team budget enforcement
    \item Expenditure tracking
    \item Personally identifiable information (PII) masking via Microsoft Presidio
    \item Comprehensive user instrumentation to persistent storage (PostgreSQL)
\end{itemize}
 
MLflow integration provides model evaluation and observability primitives that downstream consumers use to analyze usage and improve reliability and cost-efficiency.
 
\subsection{Application Augmentation Layer: RSE-Plugins Ecosystem}
\label{subsec:plugins}
 
The application layer implements domain-specific research workflows through RSE-Plugins, an ecosystem designed for research software engineering (RSE) and scientific computing tasks. Rather than relying on generic LLM assistance, RSE-Plugins provide Claude Code with specialized knowledge modules that encode established practices from the scientific Python community and the professional RSE discipline. The ecosystem is organized as a three-level Plugin-Agent-Skill hierarchy of installable, shareable packages, in which each level supplies progressively narrower and more situated guidance. Section~\ref{sec:rse_plugins} details this internal structure and the four plugins currently maintained by our team.

%% file: content/4_rse_plugins.tex
Building on the hierarchy introduced in Section~\ref{subsec:plugins}, an \emph{Agent} is a domain-expert persona that drives multi-step interactions, a \emph{Skill} is a focused, reusable knowledge module invoked on demand, and \emph{slash commands} provide explicit entry points to recurring workflows. Plugins bundle these components into installable packages that can be shared across projects, teams, and institutions. Rather than closing the gap between general-purpose coding agents and scientific Python practice through retraining or fine-tuning, this approach encodes accumulated RSE craft as composable context that any sufficiently capable agent can consume at inference time. The design philosophy follows the Scientific Python Development Guide: collaborate with community conventions rather than work around them, refactor confidently under the protection of comprehensive testing, and prefer wide, reusable solutions over deep, monolithic ones.
 
\subsection{Scientific Python Development Plugin}
\label{subsec:python_plugin}
 
The Scientific Python Development Plugin is the foundational layer, encoding the conventions and tooling on which the scientific Python community has converged over the past decade. Unlike generic Python guidance, which tends to optimize for web and application idioms, this plugin treats reproducibility, numerical correctness, and packaging discipline as first-class concerns.

\subsubsection{Scientific Python Expert Agent}

Drawing on the Scientific Python Development Guide, this agent offers guidance across the core scientific Python stack (NumPy, Pandas, Matplotlib, and SciPy); package architecture using src-layout conventions, \texttt{pyproject.toml}, and the Hatchling build backend; reproducible environment management with \texttt{pixi} (unified Conda and PyPI resolution with multi-platform lockfiles); testing with \texttt{pytest}, NumPy testing utilities for numerical comparisons, and Hypothesis for property-based testing; code quality tooling such as \texttt{ruff} for linting and formatting, \texttt{mypy} for static type checking, and \texttt{pre-commit} hooks; and numerical computing best practices including edge case handling (NaN, inf, empty arrays), separation of I/O from computation logic, and NumPy-style docstrings. Rather than dispensing isolated point recommendations, the agent applies a structured decision-making framework that weighs scientific context, reproducibility requirements, and community conventions when guiding implementation choices.

\subsubsection{Scientific Documentation Architect Agent}

Documentation quality is critical to scientific software discovery, adoption, and long-term maintenance, yet documentation is consistently the work researchers most often defer. This agent produces technical documentation for scientific Python codebases, organized according to the Di\'ataxis framework. Core competencies include Di\'ataxis-structured documentation (tutorials, task-oriented how-to guides, API reference, and conceptual or architectural explanation); Sphinx and MkDocs configuration with \texttt{pydata-sphinx-theme}, \texttt{furo}, \texttt{numpydoc}, \texttt{autodoc}, \texttt{napoleon}, and \texttt{intersphinx} cross-referencing; NumPy-style docstrings and API reference generation for scientific libraries; visual communication through architecture diagrams and algorithm flowcharts; and multi-audience technical writing for researchers, developers, and domain experts. The agent proceeds through a four-phase workflow (Discovery, Planning, Structuring, and Writing) intended to produce documentation that is complete, reproducible, and accessible across experience levels.

\subsubsection{Scientific Python Skills}

Five specialized skills support the plugin's agents, each encoding a focused slice of scientific Python practice:

\paragraph{\texttt{pixi-package-manager}:} Managing dependencies with \texttt{pixi}, which unifies Conda and PyPI in a single resolver. Covers multi-platform lockfiles (Linux, macOS, Windows), feature-based environment configurations (development, testing, GPU/CPU), and task automation.

\paragraph{\texttt{python-packaging}:} Modern packaging conventions: src-layout, \texttt{pyproject.toml}, the Hatchling build backend, CLI entry points, PyPI distribution, and version and dependency specification appropriate to scientific packages.

\paragraph{\texttt{python-testing}:} \texttt{pytest}-based testing for scientific code: NumPy numerical comparisons with floating-point tolerances, fixtures and parametrization, property-based testing with Hypothesis, coverage measurement, and CI setup.

\paragraph{\texttt{code-quality-tools}:} \texttt{ruff} for unified linting and formatting (replacing \texttt{flake8}, \texttt{black}, and \texttt{isort}), \texttt{mypy} for static type checking, and \texttt{pre-commit} hooks for automated quality gates, with CI integration.

\paragraph{\texttt{scientific-documentation}:} Sphinx and MkDocs configuration, NumPy-style docstrings, the Di\'ataxis framework, accessibility standards, and documentation hosting on Read the Docs.
 
\subsection{Scientific Domain Applications Plugin}
\label{subsec:domain_plugin}
 
Where the Scientific Python plugin captures cross-cutting practice, the Scientific Domain Applications Plugin addresses the data formats, conventions, and computational patterns that distinguish specific research communities. Astronomy and Earth/climate science are the first instantiations, reflecting where our center has accumulated the deepest engagement; agriculture, health, and additional geoscience subfields named in our portfolio (Section~\ref{sec:introduction}) are an active workstream and we are actively working on representing these as dedicated plugins.

\subsubsection{Astronomy \& Astrophysics Expert Agent}

This agent consolidates expertise in astronomical computing workflows (\texttt{FITS} file handling, celestial coordinate transformations, photometric and spectroscopic pipelines, and physical unit management) anchored in the Astropy ecosystem. It guides users through workflows in which heterogeneous data types and successive transformations would otherwise invite errors~\cite{josephAstroVisBenchCodeBenchmark2025}.

\subsubsection{Scientific Domain Skills}

Two focused skills support domain-specific expertise:

\paragraph{\texttt{xarray-for-multidimensional-data}:} Labeled multidimensional arrays with Xarray: \texttt{NetCDF}, \texttt{HDF5}, and \texttt{Zarr} I/O; Dask integration for out-of-core datasets; DataTree for hierarchical organization; and \texttt{rioxarray} for geospatial rasters. Particularly valuable for climate, Earth science, and satellite workflows.

\paragraph{\texttt{astropy-fundamentals}:} \texttt{FITS} I/O, celestial coordinate transformations and catalog cross-matching, physical units and quantities, multi-scale time handling, aperture and PSF photometry with \texttt{photutils}, and 1D spectroscopy with \texttt{specutils}.

\subsection{AI Research Workflows Plugin}
\label{subsec:workflow_plugin}
 
Unstructured AI assistance has a documented tendency to produce bloat, undisciplined refactoring, and implementation approaches that are difficult to justify after the fact~\cite{labanLLMsGetLost2025}. The AI Research Workflows Plugin imposes a structured methodology for complex software tasks, with explicit decision-making at each stage and auditable trails of technical reasoning emitted as a routine byproduct.
 
\subsubsection{Research Workflow Orchestrator Agent}
 
This agent guides users through a rigorous six-phase development methodology in which each phase produces durable artifacts consumed by subsequent phases:
 
\begin{enumerate}
    \item \textbf{Research Phase} (\texttt{/research} command): Systematic documentation of existing code, architectural patterns, and dependencies. Findings are saved as structured markdown to the \texttt{.agents/} directory for reference in subsequent phases. This enables a spec-driven development ~\cite{piskala_spec-driven_2026} workflow where specifications become executable, directly generating working implementations. 
    
    \item \textbf{Planning Phase} (\texttt{/plan} command): Creation of detailed, testable implementation plans through interactive research. The agent decomposes complex tasks into phased units with measurable success criteria split into automated and manual checks.
    
    \item \textbf{Plan Iteration} (\texttt{/iterate-plan} command): Refinement of plans based on user feedback or changed requirements, maintaining plan consistency and completeness while accommodating evolving understanding.
    
    \item \textbf{Experimentation Phase} (\texttt{/experiment} command): Optional exploration of multiple implementation approaches before commitment, enabling informed selection of optimal strategies.
    
    \item \textbf{Implementation Phase} (\texttt{/implement} command): Execution of the plan with systematic verification checkpoints. Implementation proceeds phase-by-phase with intermediate validation against plan specifications.
    
    \item \textbf{Validation Phase} (\texttt{/validate} command): Systematic verification that implementation satisfies plan criteria by running all automated checks defined in the plan, identifying required manual testing steps, and generating a structured validation report.
\end{enumerate}
 
\subsubsection{Research Workflow Skills}

Supporting the orchestrator agent, the plugin provides one specialized skill:

\paragraph{\texttt{research-workflow-management}:} Templates and frameworks supporting the six-phase methodology, including research documentation templates capturing domain analysis and architectural decisions; plan templates specifying implementation phases, measurable success criteria, and file-level references; experiment templates for controlled comparison of implementation approaches; implementation templates with per-phase verification checkpoints; and handoff templates supporting knowledge transfer between collaborators and workflow sessions.
 
By documenting decisions at each step of research, planning, and implementation, the workflow addresses a critical limitation of unstructured AI assistance---the absence of decision transparency---and extends scientific reproducibility from data and code to the software engineering decisions themselves. Section~\ref{sec:capabilities} situates this workflow alongside the other capability surfaces of LLMoxie.

\subsection{Project Management Plugin}
\label{subsec:project_plugin}

The Project Management Plugin addresses the full lifecycle of research software, from inception through mature community development and eventual transition to new maintainers---a recurring need in academic settings where contributors graduate, move institutions, or rotate off grants.

\subsubsection{Project Onboarding Specialist Agent}

This agent handles project initialization, contributor onboarding, and knowledge transfer for open-source projects in any language. It scaffolds community health files (\texttt{README}, \texttt{CONTRIBUTING}, \texttt{LICENSE}, \texttt{CODE\_OF\_CONDUCT}, \texttt{SECURITY}, \texttt{CITATION.cff}), GitHub issue and pull request templates, and onboarding documentation. For handoff, it documents institutional knowledge, audits open issues and experimental branches, and assembles transfer packages for incoming maintainers.

\subsubsection{Documentation Validator Agent}

This agent audits documentation quality and validates setup instructions, combining automated tooling with manual, step-by-step tracing of setup instructions. The automated tooling includes Vale for prose, \texttt{markdownlint}, HTMLProofer and \texttt{lychee} for links, \texttt{doc8} for reStructuredText, language-specific doctest runners such as \texttt{pytest --doctest-glob} and \texttt{cargo test --doc}, and \texttt{nbval} for Jupyter notebooks. Findings are categorized by severity (critical, important, recommended, minor) and reported with file and line references.
 
\subsubsection{Project Lifecycle Commands}

Three slash commands provide explicit entry points to recurring project-lifecycle workflows:

\paragraph{\texttt{/setup-project}:} Scaffolds new projects with community health file templates, standard directory structure, and development tooling configuration.
 
\paragraph{\texttt{/project-handoff}:} Assesses project readiness for transition to new maintainers, evaluating documentation completeness, contributor experience, and knowledge transfer requirements.
 
\paragraph{\texttt{/validate-project-handoff}:} Systematically tests that setup instructions and documentation function as written, catching discrepancies between documented and actual processes.

\subsubsection{Project Management Skills}

Two skills provide reusable knowledge modules for the plugin's agents:

\paragraph{\texttt{community-health-files}:} Templates for \texttt{README}, \texttt{CONTRIBUTING}, \texttt{LICENSE}, \texttt{CODE\_OF\_CONDUCT}, \texttt{SECURITY}, \texttt{CITATION.cff}, GitHub issue and pull request templates, and changelogs, with conventions appropriate for academic and research contexts.

\paragraph{\texttt{documentation-validation}:} Vale prose linting, \texttt{markdownlint}, HTMLProofer link checking, \texttt{pytest} doctest for code examples, container-based validation of setup instructions, and CI integration for automated documentation checks.

%% file: content/5_research_capabilities_and_operational_modes.tex
LLMoxie exposes two complementary capability surfaces, each addressing a distinct stratum of the research software lifecycle and each adoptable independently. Where unified ``AI research platforms'' commonly conflate engineering assistance and infrastructure provisioning into a single stack, we have kept these concerns orthogonal: researchers can adopt the RSE-Plugins ecosystem with any sufficiently capable coding agent, deploy the Azure substrate as a self-hosted inference and governance plane, or combine both. The remainder of this section details these two capability surfaces and the software engineering practices that underpin both.

\subsection{Engineering Assistance via RSE-Plugins}

The RSE-Plugins ecosystem (Section~\ref{sec:rse_plugins}) supplies structured context to coding agents at inference time. Through the Plugin-Agent-Skill hierarchy, researchers obtain scientific Python development discipline, domain-situated guidance for fields including astronomy and earth/climate science, a structured AI research-and-implement workflow, and project lifecycle management. Unlike approaches that pursue domain alignment through retraining or fine-tuning, this capability encodes accumulated RSE craft as composable context that any sufficiently capable coding agent can consume.

\subsection{Production Deployment on Azure}
\label{sec:infrastructure}

The infrastructure layer provisions the cloud resources required to operate LLMoxie at scale and is available to teams whose primary use of LLMoxie is the RSE-Plugins ecosystem as well as to teams operating LLMoxie as a self-hosted inference and governance plane. LLMoxie provides infrastructure-as-code capabilities through Pulumi, enabling declarative specification and automated provisioning of cloud-native resources on Azure. The deployment system creates a comprehensive infrastructure stack, including:

\begin{figure}
    \centering
  \includegraphics[width=\columnwidth]{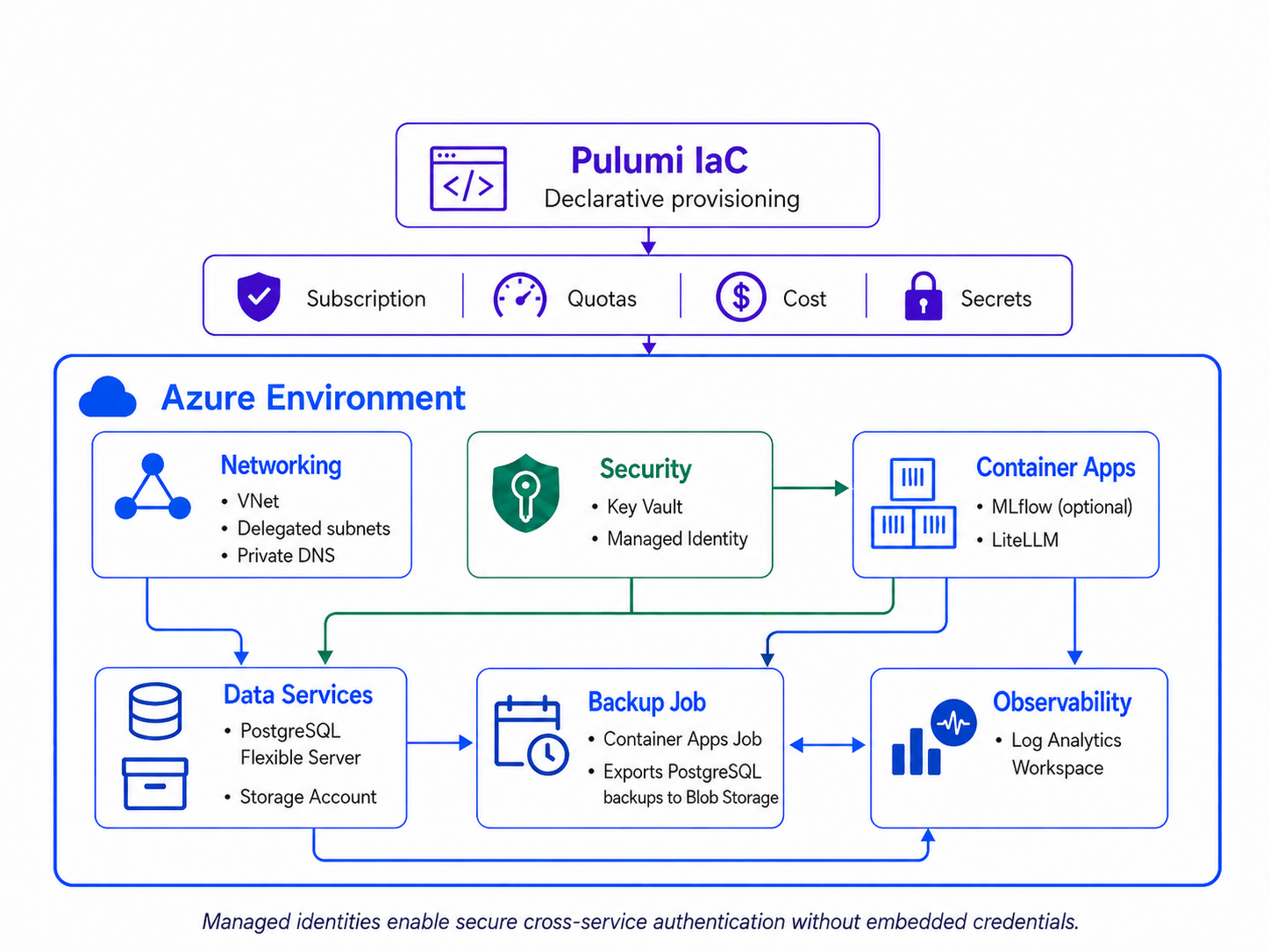}
  \caption{LLMoxie Azure deployment architecture. Pulumi declaratively provisions a virtual network with delegated subnets and private DNS, an Azure Key Vault for secrets and managed identities, a PostgreSQL Flexible Server with scheduled backups to blob storage, and Azure Container Apps hosting the LiteLLM proxy and MLflow experiment tracking, all instrumented through a Log Analytics workspace.}
  \Description{Architecture diagram of the self-hosted LLMoxie deployment on Microsoft Azure. A virtual network contains delegated subnets for Azure Container Apps and an Azure Database for PostgreSQL Flexible Server, connected through a private DNS zone. Azure Key Vault stores secrets, including the auto-generated PostgreSQL administrator password and derived connection strings, and managed identities authenticate services without exposed credentials. Container Apps host the LiteLLM inference proxy and an optional MLflow tracking server, while a scheduled Container Apps Job exports database backups to an Azure Storage Account blob container. A Log Analytics workspace collects logs and metrics across all components. The entire stack is provisioned declaratively with Pulumi.}
  \label{fig:llmoxie-cloud}
\end{figure}

\begin{itemize}
    \item \textbf{Virtual networking infrastructure} with delegated subnets for Container Apps and PostgreSQL, and a private DNS zone for database connectivity

    \item \textbf{Cryptographic key management} via Azure Key Vault with access-policy-based authorization, coupled with a secrets workflow that generates the PostgreSQL administrator password and persists derived connection strings as managed secrets

    \item \textbf{Relational database services} providing PostgreSQL Flexible Server instances with configurable resource tiers, configurable backup retention, and optional zone-redundant high availability

    \item \textbf{Object storage} via an Azure Storage Account with blob containers used for database backups

    \item \textbf{Container orchestration} via Azure Container Apps with optional support for MLflow experiment tracking and LiteLLM proxy services

    \item \textbf{Scheduled database backups} implemented as an Azure Container Apps Job that exports PostgreSQL databases to blob storage on a recurring schedule

    \item \textbf{Observability infrastructure} including Log Analytics Workspace for monitoring and diagnostics

    \item \textbf{Managed identities} for secure cross-service authentication without credential exposure
\end{itemize}

The deployment system includes validation workflows that verify Azure subscription access, validate resource quotas, estimate operational costs, and perform secrets validation before infrastructure provisioning. 

\subsection{Software Engineering Practices and Quality Assurance}

\begin{figure}
    \centering
  \includegraphics[scale=0.6]{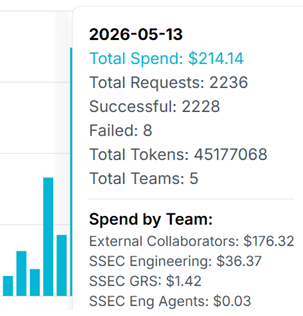}
  \caption{LLMoxie Analytics snapshot. LLMoxie platform spend for 2026-05-13, showing 2,236 API requests (99.6\% success rate), 45.2M tokens consumed, and a total spend of \$214.14 distributed across five teams, with External Collaborators comprising the dominant share (\$176.32).}
  \Description{LLMoxie platform usage daily snapshot. One day of usage for one deployed instance of LLMoxie showing 2,236 API requests (99.6\% success rate), 45.2M tokens consumed, and a total spend of \$214.14 distributed across five teams, with External Collaborators comprising the dominant share (\$176.32).}
  \label{fig:llmoxiespend}
\end{figure}

Both LLMoxie and RSE-Plugins adhere to open-source software engineering standards~\cite{hunter-zinckTenSimpleRules2021, bridgefordTenSimpleRules2025} appropriate for research software. Automated testing with \texttt{pytest} and coverage analysis is paired with \texttt{pre-commit} hooks that enforce code quality (\texttt{flake8} for linting, \texttt{prettier} for formatting consistency, \texttt{codespell} for documentation accuracy). Environments are managed through \texttt{pixi}, enabling consistent development, testing, and deployment across heterogeneous host systems while integrating both Conda and PyPI package ecosystems. Comprehensive logging and monitoring support diagnosis of production issues, with PostgreSQL integration providing persistent audit trails. Figure~\ref{fig:llmoxiespend} shows a high level snapshot of daily usage. Further drill down reports show detailed usage by provider, model and key. 
Documentation is generated automatically from source code---OpenAPI schemas for REST APIs and docstring extraction for Python modules---reducing documentation drift.

%% file: content/6_implications_and_future_work.tex
Several lessons emerge from our practice of incorporating LLMs and agents with PIs across a dozen research teams. In addition to creating and using LLMoxie for scientific software development within the academic research and open-source/open-science ecosystem, we observed measurable gains in our capacity for impact.

\textbf{Invest early in onboarding and documentation.}
Initial training sessions revealed substantial gaps in user understanding of AI-assisted scientific software development. Subsequent guided exposure helped our collaborators develop more accurate mental models of what AI, specifically the LLMoxie infrastructure and plugins, could and could not reliably do. Documentation quality proved to be a primary lever: clear, detailed documentation measurably reduced errors and improved adoption rates across teams.

\textbf{Prioritize modular infrastructure design.}
Modularity (e.g., reusable skills and plugins) allowed teams to integrate LLMoxie plugins and skills into diverse workflows as reusable, domain-specific components rather than conforming to a single prescribed approach. Model flexibility, interoperability, and an openly documented design improved debuggability and made teams more likely to adopt and sustain agentic-AI-generated code, and in several cases to contribute community plugins back to the LLMoxie open-source project.

\textbf{Calibrate best practices to user expertise.}
Greenfield development tasks showed clear and consistent gains, with LLMoxie accelerating prototyping and enabling non-experts to implement complex functionality. Complex debugging, legacy codebase modification, and tasks requiring fine-grained correctness presented a more complicated picture. Students without strong domain knowledge who used agentic AI through LLMoxie sometimes unknowingly incorporated errors into their code, echoing broader findings that novice users are most susceptible to over-reliance and to skill-formation deficits when AI does the heavy lifting~\cite{obrienThreatsScientificSoftware2025, kabirStackOverflowObsolete2024, shenHowAIImpacts2026, wuHumangenerativeAICollaboration2025}. Large undifferentiated context windows degraded performance, consistent with reports that long-context retrieval and multi-turn agent interactions exhibit characteristic degradation modes~\cite{labanLLMsGetLost2025, LongContextRAGMon08/12/2024-12:46}; preprocessing and selectively curating relevant information substantially improved output quality.

\textbf{Plan for cost, observability, and long-term sustainability.}
Centralized infrastructure providing unified access across multiple models substantially lowered barriers to entry, with shared resource allocations enabling broader participation, particularly among students and junior researchers. When LLMoxie infrastructure is funded through credit programs with expiration dates such as federally funded national AI compute pilots \cite{llmaven2026nairr}, system design must account for resource constraints and plan for sustainability. Observability and telemetry proved essential: our team used LiteLLM and MLflow to monitor usage, diagnose failures, and understand adoption patterns. These signals guided per-user-group configuration choices that optimized both token usage and corresponding spend.

\textbf{Address the governance gap.}
LLMoxie shifts developer focus from code generation toward higher-order evaluation and design, enabling more rapid prototyping and a greater willingness to explore and discard approaches. It also improved interaction with legacy systems by reducing the cognitive load of navigating unfamiliar code, making maintenance of complex legacy codebases more approachable. However, current agentic AI practices lack mechanisms analogous to version control or commit history for tracking the provenance of AI-assisted contributions, accumulating what has been termed \emph{intent debt}: the absence of externalized rationale that developers and AI agents alike need in order to safely evolve a system~\cite{storeyTechnicalDebtCognitive2026}. While some communities have begun to establish standards for reviewing AI-assisted software~\cite{SoftwareReviewEra2026, hosseiniOpenScienceGenerative2024}, this remains a rapidly evolving area. We have used this gap to scope future work on LLMoxie centered on provenance tracking, reproducibility, and auditability of AI-assisted contributions.

\textbf{Build feedback loops between practice and infrastructure.}
The most successful deployments followed a pattern of iterative co-evolution rather than linear tool adoption. LLMoxie infrastructure and plugins were built by our team to address the pain points encountered in day-to-day work on scientific software development. Evidence-based best practices---structured development phases, context-curation methods, and quality-assurance mechanisms---once designed as reusable components, became available to subsequent teams without requiring them to independently traverse the same learning curve. Institutional investment should therefore focus not only on tooling itself but on the feedback mechanisms that allow learned practices to be continuously refined and shared.

%% file: content/7_limitations.tex
Thorough utilization across domains, user experience levels, and use cases has revealed areas of improvement. First, the experience reported here is drawn from a single multi-domain RSE center; while the same failure modes recurred across diverse projects, generalization to other institutional settings remains to be established. Second, the operational lessons in Section~\ref{sec:implications} are based on practitioner observation and project artifacts rather than a controlled user study; we do not report quantitative measures of productivity, code quality, or scientific output, and a more formal evaluation is an obvious next step. Third, the application-augmentation layer is currently anchored on a specific coding agent (Claude Code), and although the plugin design targets portability across sufficiently capable agents, that portability has not been exercised at scale. Fourth, as an open-source framework for running, managing, and benchmarking LLM experiments and deployments, the repository has few active maintainers; substantial investment is needed to grow the contributor base, otherwise issues and pull requests risk accumulating and external users may fork rather than contribute. Finally, LLM-based software engineering is a rapidly evolving field with an accelerated pace of technological breakthroughs and model evolution~\cite{mcinnesReport2025Workshop2025}. While RSEs can keep up with and continually evolve tooling such as LLMoxie, prioritization of the engineering backlog must be thoughtfully balanced against unforeseen developments from frontier model laboratories.

%% file: content/8_conclusion.tex
LLMoxie offers a systematic approach to integrating large language models into scientific research workflows while preserving researcher autonomy, institutional data governance, and scientific reproducibility. Its three-tiered architecture separates inference, governance, and application augmentation, enabling flexible deployment across institutional contexts under uniform governance, while the Plugin-Agent-Skill hierarchy provides a scalable framework for encoding RSE expertise across diverse scientific computing challenges.

Through the open-source RSE-Plugins ecosystem, the platform shifts LLMs from generic code generators into domain-aware collaborators that respect community conventions, follow established scientific computing practices, and produce auditable trails of technical reasoning. By combining open-source development, modular design, and reusable RSE knowledge, LLMoxie provides a foundation for accessible, high-quality AI-augmented scientific software that allows researchers and RSEs to pair computational and human expertise without compromising the rigor and reproducibility that scientific work demands.

%% file: content/9_acknowledgments.tex
The authors thank the National AI Research Resource (NAIRR) Pilot program funded by the National Science Foundation (NSF) that provided cloud credits (Award 240292) using which the LLMoxie stack was created. We thank HuggingFace and Microsoft Azure for credits. In addition we thank support from Sarah Stone and the CloudBank program for helping sustain the development of LLMoxie. We thank Amazon and AWS Bedrock for credits to eScience Institute using which we utilized Anthropic models. The University of Washington Scientific Software Engineering Center (SSEC) is supported by Schmidt Sciences, as part of their Virtual Institute for Scientific Software (VISS) program. 